# Dynamics of Mortality in Protected Populations

Mark Ya. Azbel'

School of Physics and Astronomy, Tel-Aviv University,

Ramat Aviv, 69978 Tel Aviv, Israel[+];

Max-Planck-Institute für Festkorperforschung – CNRS,

F38042 Grenoble Cedex 9, France.

## Abstract

Demographic data and recent experiments verify earlier predictions that mortality has short (few percent of the life span) memory of the previous life history, may be significantly decreased, reset to its value at a much younger age, and (until certain age) eliminated. Such mortality dynamics is demonstrated to be characteristic only of evolutionary unprecedented protected populations. When conditions improve, their mortality decreases stepwise. At crossovers the rate of decrease rapidly changes. The crossovers manifest the edges of the "stairs" in the universal "ladder" of rapid mortality adjustment to changing conditions. Mortality is dominated by the established universal law which reduces it to few biologically explicit parameters and which is verified with human and fly mortality data. Specific experiments to test universality of the law for other animals, and to unravel the mechanism of stepwise life extension, are suggested.

[+]   Permanent address

Phone, fax:  +972-3-540-7874          e-mail: azbel@post.tau.ac.il

**Introduction**

Accurate knowledge of human mortality is important for economics, taxation, insurance, etc. The famous astronomer Halley, 1693 (the discover of the Halley comet) started quantitative mortality studies. He was followed by the great mathematician Euler, 1760. The actuary Gompertz, 1825 presented the first universal law of mortality for human advanced age. Thereafter the search for such law for all animals went on (Pearl, Miner, 1935; Deevey, 1947; Strehler, Mildvan, 1960; Carnes et al, 1996; Azbel, 1996, 1998) and lead to the maximal life span paradigm. (For more details see Olshansky, Carnes, 1997 and A1-here and on Appendixes are denoted by A followed by their number). To better estimate and forecast mortality, demographers dropped the universal law and developed over 15 mortality approximations (Coale et al, 1993; Lee, Carter, 1993), each of them for a given population, e.g., for the Southern Italy. Biologists discovered failure of the maximal life span paradigm (Carey et al, 1992; Curtsinger at al, 1992). Evolutionary theories of mortality were presented (Kirkwood, Austad, 2000; Finch, Kirkwood, 2000; Charlesworth, 1994). Physicists numerically tested them (Stauffer, 2002). Biogerontologists performed biologically motivated partitioning of mortality (Carnes, Olshansky, 1997) and suggested that the universal law is valid for intrinsic mortality only (Carnes et al, 1996). Yet, 180 years after Gompertz, the existence of the universal (i.e. independent of the population, its living conditions and life history) relation between certain (non-universal) mortality characteristics is still controversial. Meanwhile, accurate formulation of the problem allows for its exact mathematical solution. The resulting universal relation (Azbel, 2002a, 2003) reduces total mortality of any population at any age to its infant mortality in the same calendar year and several other biologically explicit parameters. With few percent accuracy it approximates male and female mortality curves in 18 developed countries over their entire history (except the years during, and immediately after, wars, epidemics, etc) according to Human Mortality Database, 2003 (further HMD). Universal data fitting of over 3000 curves with only few adjustable parameters (A2) proves that it presents the exact universal law. Properly scaled, the law is valid for flies also (Azbel, 1998, 1999, 2002b). It implies that 1) At any age mortality is as plastic as "infant" (till 1 year for humans, 1 day for flies) mortality, has short memory of the previous life history, and rapidly (within few percent of the life span) adjusts to current living conditions; 2)At least until certain age (80 years for humans) it may be significantly decreased and even eliminated; 3)Mortality of a homogeneous cohort may be reset to its value at a much younger age.

Demographic data and recent experiments verify all these conclusions. 1) Following unification of East and West Germany, within few years mortality in the East declined toward its levels in the West, especially among elderly, despite 45 years of their different life histories (Vaupel et al, 2003). Dietary restriction resulted in essentially the same robust increase in longevity in rats (Yu et al, 1985) and decrease in mortality in Drosophilas (Mair et al, 2003), whenever it was switched on, i.e. independent of the previous "dietary" life history. 2) Arguably, zero mortality was also observed, but not appreciated. In 2001 Switzerland (HMD) only 1 (out of 60,000) girl died at 5, 9, and 10 years; 5 girls died in each age group from 4 to 7 and from 9 to 13 years; 10 or less from 2 till 17 years; no more than 16 from 2 till 26. Statistics is similar in all 1999-2002

Western developed countries (HMD). Such low values of a stochastic quantity strongly suggest its zero value, at least in lower mortality groups. Similarly, only 2 (out of 7500) dietary restricted flies died at 8 days (Mair et al, 2003). Jazwinski et al, 1998 presented the first model which stated that a sufficient augmentation of aging process resulted in a lack of aging, and supported this conclusion with experimental evidence on yeast mortality. Changes in small number of genes and tissues in nematods C. Elegans (Arantes-Oliveira et al, 2003) increased their mean life span six-fold to 124 days (compared to 20 days of the wild type) with no apparent loss in health and vitality. None of these 1368 nematods died till 25 days, only 5% died till 40 days (twice the wild life span), and only 15% died in the first 3 months. Female survivability in 1999 Japan was (HMD) 97% till 51 year, 95% till 58 y, 85% till 73 y, 73% till 80 y (cf 6% in 1950). 3) Mortality of the female cohort, born in 1900 in neutral Norway, decreased after 17 years of age, at 40 reversed to its value at 12 years; then little changed till 50 years, and only at 59 years restored its value at 17 years, i.e. 42 years younger. (The cohort probability $q(x)$ to die at any age x is calculated according to the HMD procedures and data. According to the universal law, such mortality decrease, similar to the Germanys unification, is not dominated by death of the frail which alters the composition of the cohort-see Discussion). Dietary restriction, switched on day 14 (Mair et al, 2003), in 3 days restored Drosophila mortality at 7 days, i.e. 10 days younger.

Thus, under certain conditions, predicted short memory and reversal of mortality to much younger age are indeed universally observed in flies, rats, and humans; vanishing and very low mortality is universally seen in yeast, nematods (including biologically amended ones), flies, and humans. However, in some cases (e.g., when diet is switched off) this is not true. Presented study specifies conditions when the universal law is valid, and suggests experiments which may comprehensively verify it and its implications. Similar experiments may also establish biological limits of universality. Validity of the universal law implies that its mortality till certain age may be eliminated, and life span correspondingly extended, while universality of the (properly scaled) law implies that its mechanism is common to different species, genotypes, phenotypes, living conditions, life histories and all other factors. So, its most primitive biological level allows for the simplest case study of its mechanism (for instance, universality in yeast would reduce it to processes in a cell) and the ways to direct it to life extension. (More biological details see in Discussion).

## Materials and Methods

Demographic life tables present millions of accumulated mortality data in many countries over their history (see, e.g., HMD), which increased infant mortality 50-fold and doubled the life expectancy (A1). For males and females, who died in a given country in a given calendar year, the data list, in particular, "period" probabilities q (x) (for survivors to x) and d(x) (for live newborns) to die between the ages x and (x+1) [note that d(0)=q(0)]; the probability $l(x)$ to survive to x for live newborns; the life expectancy e(x) at the age x for males and females who died in a given country in a given calendar year. They also present (see, e.g., HMD) the data and procedures which allow one to calculate the values of q(x), d(x), $l(x)$, e(x) for human cohorts, which were born in a given calendar year. All these data, together with fly life tables (Carey, 1993), are used as "materials". Short mortality memory, combined with strong mortality dependence on living conditions (e.g., in 20 years the probability to survive from 80 to 100 years increased 100-fold-Vaupel et al, 2003), suggest to study period data and their dependence on q(0). Mathematics provides the method (A2, see Azbel, 2003 for technical details) which allows one to unravel universality and establish its law in uncontrollable non-stationary and heterogeneous conditions, to accurately partition the relative contribution of different non-universal mechanisms, and to unravel a new unusual mechanism of universal mortality.

## Results

Mortality in protected populations of humans and flies is dominated by the universal law (see A2 for details on this section). It is valid only in evolutionary unprecedented well protected populations, when "infant" (till 1 year for humans, 1 day for flies) mortality q(0) is small (less than 0.15 for humans-this specifies a protected population) and, e.g., for humans relatively little changes from one calendar year to the next (this defines "regular" conditions, in contrast to "irregular" ones during, and immediately after, wars, epidemics, etc). During the life span of a generation infant mortality change may be very large (~50-fold), i.e. very rapid on such time scale.

Start with the special case when all values of infant mortality in different groups of a given population belong to only one (out of 5) interval between two specified universal constants (e. g., between 0.011 and 0.025 for humans). Then the universal law reduces the total period mortality of such population (further denoted as an "echelon") to its infant mortality only. The law is most explicit (Azbel, 2003) for the period probability d(x) for a live newborn to survive to a given age x: it linearly depends on the infant mortality only, it is independent of the distribution of specific values of the infant mortalities in the echelon, as well as of different genotypes, phenotypes, conditions, life histories, which lead to these values. In contrast, the law is different in different echelons and rapidly changes at their boundaries.

In a general case different groups in the population (e.g., rural/urban, rich/poor, educated/uneducated human ones) belong to different echelons. Then the population period mortality $q(x)$ at any age universally depends on the infant mortalities and the fractions of the population in these echelons. In regular conditions the period probability $d(x)$ linearly changes with $q(0)$ until infant mortality in one of the echelons reaches its boundary. Then the rate of this change switches to a different value. Such piecewise linear dependence is exemplified in figures 1 and 2 for females in countries as different as Japan and France. The crossovers at $q(0) \sim 0.006$ amplify significant declines of old age mortality in the second half of the 20-th century (Wilmoth, Horiuch, 1999; Wilmoth et al, 2000; Tuljapurkar et al, 2000) with its spectacular medical progress. The universal law allows one (with few percent accuracy) to approximate $d(x)$, and thus $q(x)$, with several age independent adjustable parameters (Azbel, 2003, A2). The approximations for both sexes in all 14 HMD developed countries for all but 'irregular" years reduce to 5 universal echelons, with rapid change of the linear $d(x)$ vs $q(0)$ dependence at their boundaries (A3). When infant mortality vanishes, the universal law suggests vanishing human mortality at any age till ~80 years (in agreement with human mortality data in the Introduction), its universal minimal values in older age, and life expectancy at birth ~97 years (in agreement with Azbel, 1996).

Infant mortality depends only on short period of time (for humans, e.g., less than 2 years, from conception to 1 year). So, universal mortality at any age has correspondingly short memory. This implies its correspondingly rapid adjustment to changing living conditions, and a possibility of its reversal to much younger age.

The universal minimal and maximal mortality values in each echelon demonstrate and quantify the boundaries of the "stairs" in the universal "ladder" of mortality dynamics in progressively improving protected conditions (A3). Flies yield the same (when properly scaled) universal law (Azbel, 1999, 2003). The universal law which is preserved in evolution of species as biologically remote as humans and flies, is arguably a conservation law in biology and evolution (Azbel, 2003).

Consider biological and demographic implications of the universal law and their challenges.

**Discussion**

A crucial implication of the universal law is its plasticity, which is closely related to its short memory. The latter is very explicit in experiments where dietary restriction in rats and flies is switched on (Yu et al, 1985, Mair et al, 2003). However, when dietary restriction changes to full feeding, their longevity remains higher than in the control group of animals fully fed throughout life. Also, when fly temperature was lowered from 27 to 18 degrees or vice versa, the change in mortality, driven by life at previous temperature, persisted in the switched flies compared to the control ones. Such long memory of the life history may be related to rapid changes in temperature or feeding (mimetics included), since universal law is valid when infant mortality little changes within a day for flies, a month for rats, a year for humans (but may significantly change within their life span). This calls for comprehensive tests of short mortality memory in, and thus of rapid compared to life span mortality adaptation to, such conditions. Similar tests may verify a possibility to reverse and reset mortality of a homogeneous cohort to a much younger age.

Human universal mortality law is comprehensively verified with extensive demographic data in very different and continuously changing conditions. An accurate criterion establishes that the population in a developed country at any time predominantly reduces to two echelons (see A3 for details on this section). Certain groups in a population may belong to a single echelon, and simplify a comprehensive study of universality.

Animal life tables refer to a limited number of populations (which may include pure lines) with rather low statistics. However, reduction of any population to few echelons allows for quantitative study of their universality. Divide a given population into several groups in the same conditions. Quantify statistical differences between the group mortalities until a fixed age. Then slowly implement different conditions in different groups to increase their mortality differences significantly beyond statistical ones. Thereafter their mortalities allow one to estimate the number of co-existing echelons in these groups. The lower this number is, the less groups suffice to establish the universal law (A2). The most biologically primitive animals which yield universality may elucidate its mechanism and the nature of "adjustment stairs" in the universal "mortality ladder". If, e.g., slow changes to different temperatures or erythromycin concentrations demonstrate short memory and universality in yeast mortality, then universality may be reduced to processes in a cell.

The universal law is independent of the population, its life history and living conditions. So, it must be related to accurate intrinsic response to their change. Thus, within its accuracy total mortality of a well protected population equals intrinsic mortality (this specifies the Carnes, Olshansky, 1997 suggestion that the universal law is valid for intrinsic mortality only), while extrinsic mortality is negligible.

Universal mortality q(x) at any age x is related to infant mortality q(0). Thus, it rapidly adjusts to, and is determined by, current (<2 years for humans) living conditions only. It is independent of the previous life history and, together with q(0), it may be rapidly reduced and reversed to its value at a much younger age. So, when mortality of a cohort is predominantly regular, it may be reversed also. This agrees with Fig. 3 and mortality data in the Introduction, and calls for a comprehensive test.

Non-zero minimal universal mortality beyond certain age implies universal maximal mean life span (whose upper limit is the maximal individual life span). For humans the universal law extrapolates it to ~97 years (Azbel, 2003). This agrees with human maximal lifespan, which remains ~120 years since ancient Rome (where birth and death data were mandatory on the tombstones) to present time. This is consistent with Strehler, Mildvan, 1960; Azbel, 1996; Carnes et al, 2003.

The law which is universal for species as remote as humans and flies is presumably valid for biologically slightly amended animals, like nematods with 6-fold (compared to wild type) mean life span increase due to evolutionary unmotivated, and unprecedented, changes (Arantes-Oliveira et al, 2003). Their very low mortality in extremely old age (see Introduction) suggests a possibility of high maximal life span plasticity to such, relatively minor, amendments. This is consistent with Oeppen, Vaupel, 2000; Vaupel et al, 2003. [Note, however, similar effects in non-amended flies (Carey et al, 1992; Curtsinger et al, 1992; Carey, 1993), possibly related to significant genetic variations- Azbel, 1996; Carnes, Olshansky, 2001 and refs. therein. Note also that human mean life expectancy more than doubled in just a century].

Consider the implications of the universal law for evolutionary theories of mortality (Kirkwood, Austad, 2002; Finch, Kirkwood, 2000; Charlesworth, 1994). Mortality is an instrument of natural selection. Survival of the fittest and death of the frail depend (besides genetics) on the previous diseases, traumas, natural disasters, etc, i.e. on the entire life history. This is inconsistent with short mortality memory. In the wild, competition for sparse resources is fierce, and only relatively few genetically fittest animals survive to their evolutionary "goal"- reproduction. Even human life expectancy at birth was around 40-45 years just a century ago (e.g., 38.64 years for males in 1876 Switzerland-HMD). Thus, there are no evolutionary benefits from genetically programmed death or longevity of very few survivors to older age. The Kirkwood disposable soma theory relates their mortality to life-history trade-off (in optimal allocation of metabolic resources between somatic maintenance and reproduction). Thus, it also implies life-history mortality dependence, which is inconsistent with short memory. Theories of irreparable cumulative damage (to DNA, cells, tissues and organs) relate mortality to accumulating mutations with late-acting deleterious effects; telomeres; free radicals etc. In a homogeneous cohort such damage implies persistent mortality increase with age, which is inconsistent with mortality reversal and reset to much younger age. Significant mortality decrease and possibly elimination until old age imply that until such age (extremely old in slightly amended nematods) cumulative damage is either negligible or repairable. According to the Williams antagonistic pleiotropy theory, genes with good early effects may be favored by selection, although these genes had bad effects, including senescence and death, at later ages. In pre-reproductive and reproductive age, these genes are beneficial for longevity. Thus, in perfect conditions with abundant resources they do not yield mortality, in agreement with the universal law and experimental data in the Introduction. Finite mortality prior to post-reproductive age may be consistent with a different kind of antagonistic pleiotropy. Genes, which are beneficial for longevity in the wild, may be detrimental for universal mortality, which dominates in evolutionary unprecedented protected conditions but was negligible in the wild. Some genes, related to certain biological characteristics, may at any age be (accidentally) beneficial for longevity ("longevity genes"-see Jazwinsky, 1996; Puca et al, 2001; Atzmon et al, 2003; Nawrot et al, 2004). However, universal law implies an accurate, reversible, rapid, stepwise, intrinsic response of the total mortality to current environment. The response transforms all multiple environmental parameters into the infant mortality only. It is independent of genotypes, phenotypes, their conditions and life history. It suggests a possibility of mortality elimination (till certain age). Such law is hardly consistent with any genes. Together with its other implications, it suggests a new mechanism of mortality in protected populations. Arguably, mortality dynamics manifests a new kind of law which does not reduce to known science in the same way as statistical mechanics of any system does not reduce to mechanics of its constituent particles: the former is always reversible, the latter is not (hence the well known entropy increase).

Similar to mortality, aging beyond reproductive age yields no evolutionary benefits, and may also be related, e.g., to irreparable cumulative damage. However, biologically amended nematods challenge the inevitability of aging also. Arantes-Oliveira et al, 2003, addressed aging by examining level of activity of surviving animals. Their dynamics may be quantified. This may allow one to study its correlation with, and relation to, the universal mortality.

The universal law allows for quantitative partitioning of the total mortality (which specifies biologically motivated partitioning of mortality by Carnes, Olshansky, 1997). For a given country, sex, and calendar year it determines the population fractions in the echelons. The dependence of these fractions on, e.g., life history quantifies the impact of antagonistic pleiotropy. The difference between the total mortality and the universal law may be partitioned into stochastic fluctuations (which yield the well known Gaussian distribution), "irregular" fluctuations (related to, e.g., 1918 flu pandemic in Europe and World Wars), and systematic deviations (related, in particular, to evolutionary mechanisms). Survivors to very old age are more robust, their mortality is more related to genetics, mutation accumulation and other cumulative irreparable damage (Azbel, 1996). This increases the deviation from the universal law- e.g, for Japanese females in Fig. 1 from ~2% at 60 and 80 years to ~10% at 95 years.

**Conclusions.**

A dominant mechanism of mortality changes with age and living conditions. In the wild natural selection is followed beyond certain age by the antagonistic pleiotropy (including its new kind from the previous section), life-history trade-off, and then, in older age, by mutation accumulation and other irreparable damage. This paper considers populations, whose protection from environment (elements of nature, predators, shortage of resources, etc); social and medical (biological interventions, e.g., heart transplants and artificial hearts, included) help for humans; biological changes (e.g., genetic) in animals were evolutionary unanticipated. When conditions persistently improve, mortality descends down the "mortality ladder", with the spurts at the (simultaneous for all ages) edges of its successive "stairs" (see examples in Figs. 1 and 2), which are related to a new biological concept of population "echelons". The ladder, i.e. properly scaled mortality law, is universal, its mechanism is common to species as, and even more, remote as humans and flies. Between the spurts mortality has short memory, and impact of life history is lost. This implies the possibility of mortality reversal to a much younger age, and of its rapid (compared to the life span) stepwise adjustment to changing conditions. The possibility of significant mortality decrease, and its elimination until (with relatively minor biological amendments) extremely old age, implies, in particular, that at least until such age cumulative damage may be negligible. Universal mortality law, its spurts and all other biologically implications suggest that its mechanism is common to very different animals, thus biologically non-specific, and may be universally regulated, together with mortality and even certain aspects of aging (possibly with a non-specific "longevity pill"). The previous section suggests comprehensive experimental study of the universal law and its implications. If the study demonstrates that the law is valid in yeast also, the universal mechanism of mortality may be reduced to processes in a cell, and suggest the ways to direct mortality dynamics to successive stepwise life extension.

**Appendix 1**

Total mortality depends on a multitude of factors which describe all kinds of relevant details about the population and its environment, from conception to the age of death: genotypes, life history; acquired components, even the month of death (Doblhammer,

Vaupel, 2001) and the possibility of its being the late onset genetic decease (Partridge, Gems, 2002). Age specific factors are also important. From 1851 to 1900 English female mortality decreased 2.6 times for 10 years old and by 5% for breast fed infants, prevented from contaminated food and water (Wohl, 1983). Crop failure in 1773 Sweden increased infant mortality by 30%, and 3.6 times mortality at 10-14 years, above its value in 1751 (Statistisk Arsbok, 1993). The 1918 flu pandemic in Europe accounted for 25% of all deaths, with 65% of the flu deaths at ages 15 and 40 years. It increased Swedish female mortality threefold at 28 years, but little changed it for newborns and elderly. Strongly tubercular mortality pattern in Japan prior to 1949 (Johnson, 1995) yielded equal infant mortalities in 1947 Japan and 1877 Sweden. During the last century period mortality rates in a given country at 0, 10 and 40 years decreased correspondingly 50, 100, and 10 times. In contrast to such mortality decrease (primarily due to improving living conditions, medical ones included), the difference between period mortality rates at the same age and in the same calendar year in different countries is rarely more than twofold.

To elucidate the significance of current living conditions, i.e. the relative role of nurture vs. nature in mortality, Fig. 4 presents the change in the period mortality rates in Switzerland (during 125 peaceful years, from 1876 till 2001) and in England (from 1841 till 1998, which include the Victorian 1850-1900 years of contaminated food and water, and two World Wars). Their mortality rate at 10 years decreased more than 100 fold; infant mortalities, mortalities at 40 and 80 years decreased correspondingly 50, 10 and 5 fold. In contrast to such mortality decrease (primarily due to improving living conditions, medical ones included), the difference between Swiss and English mortalities at the same age in the same calendar year is rarely more than twofold.

**Appendix 2**

In a given echelon the period probability for a live newborn to die at the age x is (Azbel, 2003)

$$d_j(x)=a_j(x)q_j(0)+b_j(x) \quad \text{if} \quad q_j < q_j(0) < q_{j+1} \tag{1}$$

where the subscript j denotes the ordinary number (1, 2,…, J) of the interval; $a_j(x)$ and $b_j(x)$ are universal functions of x; and $q_j$ is a universal constant. At any age $d_j(x)=d_{j+1}(x)$ at the crossover $q_j(0)=q_{j+1}$. This reduces all $d_j(x)$ to (J+1) universal functions of x and (J-1) universal constants.

Suppose in different echelons the fractions of populations and of its infant mortality q(0) are correspondingly $c_j$ and $f_j = c_j q_j(0)/q(0)$. Then Eq. (1) reduces $d(x)= c_1 d_1(x)+ c_2 d_2(x)+…+c_J d_J(x)$ in any given population to the universal dependence on these biologically explicit parameters and q(0):

$$d(x)=aq(0)+b; \quad a=f_1 a_1 +…+f_J a_J ; \quad b=c_1 b_1 +…+c_J b_J \tag{2}$$

where $0<c_j, f_j<1$; $c_1+…c_J= f_1+…+f_J=1$. (Here and on I skip the argument x in a and b). Since $d(x)=[\ell(x) - \ell(x+1)]$, so $q(x) = [\ell(x) - \ell(x+1)]/\ell(x)$ equals

$$q(x)=d(x)/[1-d(0)-d(1)-…-d(x-1)] \tag{3}$$

In a general case three groups are sufficient to establish the linear segment (2), more groups allow for the verification of mortality universality with age.

**Appendix 3**

Suppose a heterogeneous population reduces to two (e.g., the 1-st and 2-nd) echelons with the concentrations $c_1$ and $c_2 = 1 - c_1$ correspondingly. Then

$q(0) = c_1 q_1(0) + (1-c_1)q_2(0); \; d(x) = c_1 d_1(x) + (1-c_1)d_2(x).$

So, by Eq. (1,2), $q_1(0) = \alpha_1 q(0), q_2(0) = \alpha_2 q(0)$, where

$c_1 = (b_2 - b)/(b_2 - b_1)$, and $\alpha_1 = (a - a_2)/[c_1(a_1 - a_2)]; \; \alpha_2 = (a_1 - a)/[(1 - c_1)(a_1 - a_2)]$.

The crossover to the next non-universal segment occurs when, e.g., $q_1(0)$ reaches the intersection $q^U(0) = (b_2 - b_1)/(a_1 - a_2)$ of the first and second universal segments. Then $q_1(0) = q^U(0)$ implies $d^I(x) = a_1 q^I(x) + b_1$. (A subscript I denotes a population specific intersection of $d(x)$). So, all these intersections belong to the first universal linear segment in Eq. (1) or its extension (see Fig. 5) in the universal law. Such "intersection universality" is the criterion of the population always reducing to two universal segments, while $c_1$ and $c_2 = 1 - c_1$ determine the fractions of the populations at the first and second universal linear segments. Demographic data demonstrate that this is the case with the (country and sex specific) intersections in most developed countries (e. g., 1948–1999 Austria, 1921–1996 Canada, 1921–2000 Denmark, 1841–1898 England, 1941–2000 Finland, 1899–1897 France, 1956–1999 West Germany, 1906–1998 Italy, 1950–1999 Japan, 1950–1999 Netherlands, 1896–2000 Norway, 1861–2000 Sweden, 1876–2001 Switzerland-see examples in Fig.5). Approximate their empirical d(x) vs

d(0) with the minimal number of linear segments which yields the saturation of the relative mean squared deviation (it is mostly consistent with statistical accuracy- its estimate see later). Then the intersections determine the universal (i.e. the same for all countries, thus for all humans) law, presented in Fig. 5. A general case (when the population is distributed at more than two universal segments) is more complicated, and may refine Fig. 5, but it also reduces to the universal law and the echelon fractions. Then each non-universal linear segment for every age, sex and country provides two non-universal parameters (a and b in Eq. (2)) to be fitted with (J+1) universal functions (i.e. (J+1) universal parameters per every age) and with sex and country specific, but age independent parameters $c_j$, $f_j$. The number of parameters to be fitted is much larger than the number of fitting parameters. Mathematically such fitting is the accurate proof of universality and its law.

Figures 1 and 2 verify Eq. (2) with the examples of empirical d(80) vs. q(0) for Japanese and French females and their piecewise linear approximations. (Note that life tables present the probabilities per 100,000 live newborns rather than per 1 as in the paper and figures). Until $\simeq$ 65 years, d(x) decreases when q(0) increases. Beyond $\simeq$ 85 years, d(x) increases together with q(0). In between, d(x) exhibits a well pronounced maximum (naturally, smeared by generic fluctuations – see Figs. 1, 2). Consider the origin of such dependence on age. The number d(x) is proportional to the probability for a newborn to survive to x, and then to die before the age (x+1). When living conditions improve, the former probability increases, while the latter one decreases. In young age the probability to survive to x is close to 1, d(x) is dominated by the mortality rate, and thus monotonically decreases together with q(0). For sufficiently old age, low probability to reach x dominates. It increases with improving living conditions, i.e. with decreasing q(0), thus d(x) increases with decreasing q(0). At an intermediate age, when improving living conditions sufficiently increase survival probability, d(x) increase is replaced with its decrease. Then d(x) has a maximum at a certain value of q(0). Further study may yield the new lowest mortality echelon, which will dominate future mortality and its law, and will yield better statistics in old age. Such echelon may also yield the d(x) maximum at 95 and even more years of age.

The universal law reduces the period canonic mortality at any age to the same calendar infant mortality in the same echelon. This specifies demographic observation that infant mortality is a sensitive barometer of mortality at any age, and is consistent with clinical studies (Osmond, Barker, 2001) of human cohorts.

Period infant mortality rate q(0) is relatively high (see Fig. 4): it was close to q(80) a century ago, currently decreased 50 – fold, but only to ~ q(60)-even now ~0.4 % of live newborns die in the first year of life (HMD). Infant mortality drastically decreases to the minimal mortality at 10 years (~ 30 times a century ago and ~150 times to extremely low present value ~0.001%-see Fig. 4). During the last century mortality decrease was ~ 10; 4 and 2 times at 40, 80 and 95 years of age. Arguably, the impact of living conditions decreases for progressively robust survivors to advanced and especially old (over 90 years for humans) age, while the contribution of non-universal evolutionary mechanisms increases (cf the accuracy of the universal law in Fig. 1 for 60, 80 and 95 years).

The extrapolation of the Japanese piecewise linear dependence to q(0)=0 within its accuracy is consistent with the universal law d(80)=0 (see Fig.5), i.e. zero mortality at (and thus presumably prior to) 80 years. It is also consistent with the dependence of the period life expectancies e(0) at birth and e(80) at 80 years on the period birth mortality q(0) of Japanese females (Azbel, 2003). If nobody dies until 80, then e(0) = 80 + e(80). In fact, the values of e(0) and e(80), extrapolated to q(0)=0, are correspondingly 93 and 16 years. Thus, e(80) + 80 = 96 years is just 3% higher than e(0)=93.

To analyze the accuracy of the results quantitatively, consider the number D(x) of, e.g., Swiss female deaths at a given age x in each calendar year (Fig. 6). At 10 years it decreases from 126 in 1876 to 1 in 2001. At 80 years the number of deaths increases (together with the life expectancy) from 231 to 951. The number of deaths depends on the size of the population, e.g. in 1999 Japan at 80 years it is ~ 13,061 (see Fig. 6). As a function of age, D(x) decreases to a minimum (at 10 years), then increases to a maximum- D(89)=1,469 in 2001 Switzerland, and D(87)=18,338 in 1999 Japan, and rapidly vanishes around 100 years. According to statistics, the corresponding stochastic error is $\sim 2/D^{1/2}$. At 10 years of age it increases from ~20% in 1976 to ~200% in 2001 Switzerland and leads to large fluctuations in q(10)-see Fig. 6. At 40 years it is ~20%; at 80 years it is ~6% in Switzerland and ~2% in Japan. If demographic fluctuations in mortality are consistent with this (minimal for a stochastic quantity) generic error for a given age, denote the corresponding mortality as "regular". Otherwise, denote it as "irregular". Human mortality is irregular only during, and few years after, major wars, epidemics, food and water contamination, etc.

The universal law reduces dynamics of mortality in any (arbitrarily heterogeneous) population of species at least as remote as humans and flies to few parameters only. In contrast, even thermodynamics of arbitrarily heterogeneous liquids does not yield a universal law. Combined with other implications of the reversible universal law, this suggests that its theory calls for fundamentally new concepts and approach.

**Acknowledgments**

I am very grateful to I. Kolodnaya for assistance. Financial support from A. von Humboldt award and R. & J. Meyerhoff chair is highly appreciated.

**Figure Legends**

Fig. 1. The period probabilities for live newborn 1950-1999 Japanese females to die between 60 and 61 (squares), 80 and 81 (triangles), 95 and 96 (diamonds) years of age vs. infant mortality q(0). Their relative mean squared deviations from their piecewise linear approximations (straight lines) are correspondingly 2.4%, 2.3% and 10%.

Fig. 2. The period probability for live newborn 1898-2001 French females to die between 80 and 81 years of age vs. infant mortality q(0). Its relative mean squared deviation from its piecewise linear approximation (straight lines) is 3.7%. Irregular data for 1914-1918 and 1939-1947 years are disregarded.

Fig. 3. Mortality rates vs. age for the 1900 Norwegian female cohort. (The rates are calculated according to the HMD procedures and data).

Fig. 4. The change in mortality rates q(x) of 1841-1998 English (triangles) and 1876-2001 Swiss (diamonds) females with calendar year. From bottom right to top right: x=10 (black), 40 (empty), 0 (black), 80, 95 (empty signs). The years of 1918 flu pandemic and World Wars in Europe, 1851-1900 years of contaminated food and water in Victorian England are denoted by small signs. Note close mortality rates q(0) and q(80) untill the 20-th century.

Fig. 5. Universal law for d(80) and d(60) (upper and lower curves, thick lines) vs. q(0). Note that d(80) = 0 and d(60) =0 when q(0)=0. Diamonds and squares represent (non-universal) intersections for (from left to right) England (two successive intersections), France, Italy and Japan, Finland, Netherlands, Norway, Denmark, France, England correspondingly. Thin lines extend the universal linear segments.

Fig. 6. Number of female deaths in Switzerland in the first year of life (black diamonds), at 10 (black squares), 40 (white triangles), 80 (empty diamonds) and in Japan at 80 (empty squares) years of age for different calendar years. Large signs denote the 1918 flu pandemic year.

**Figure 1**

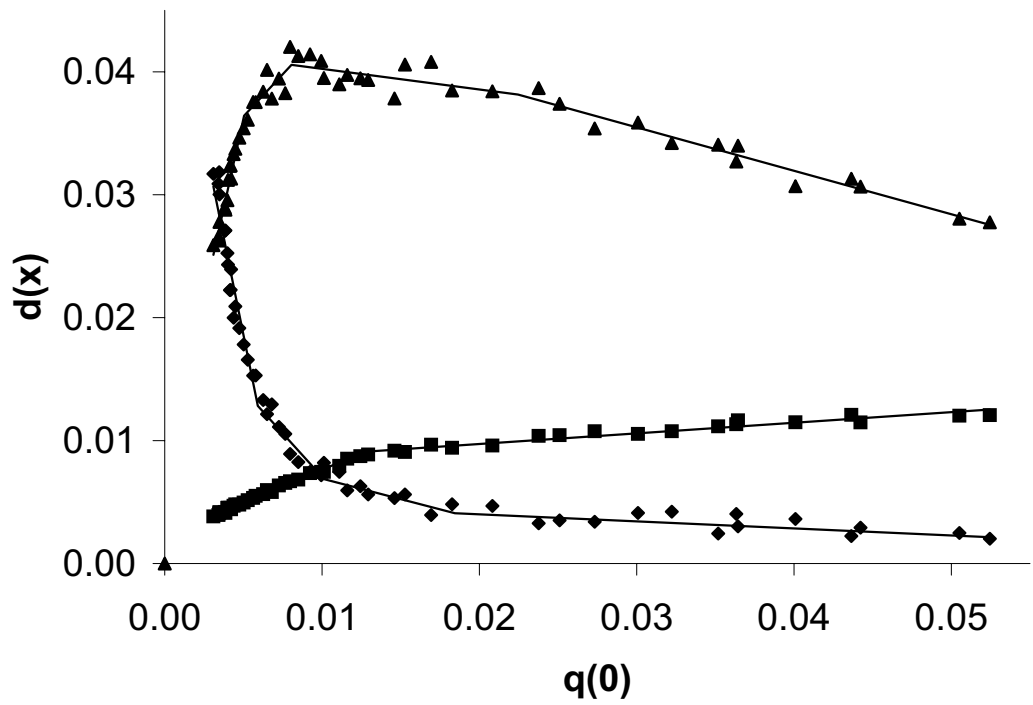

**Figure 2**

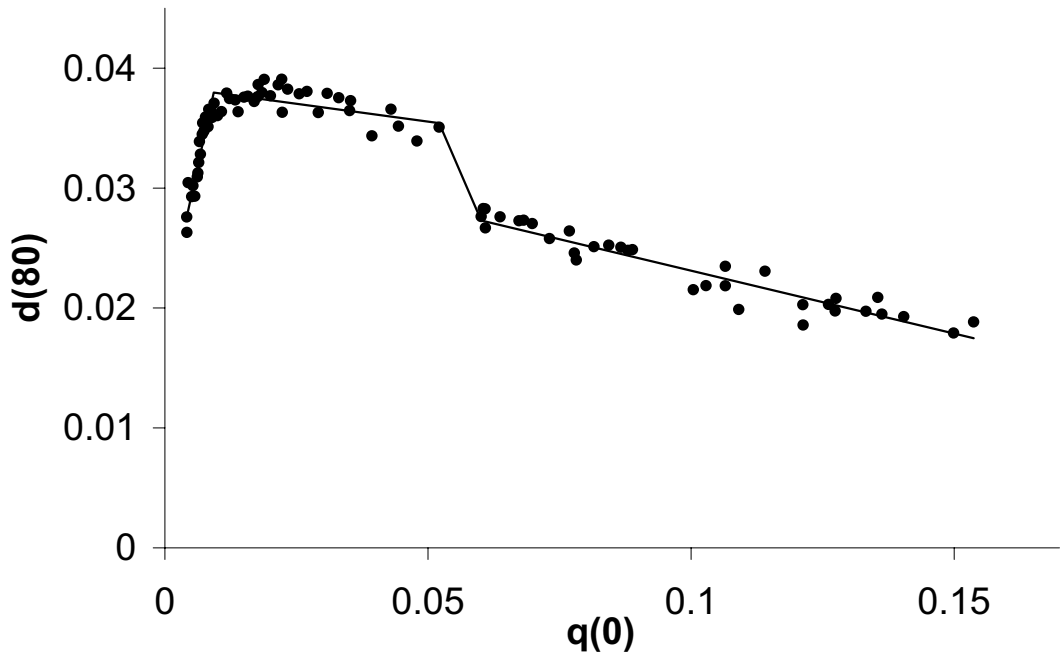

**Figure 3**

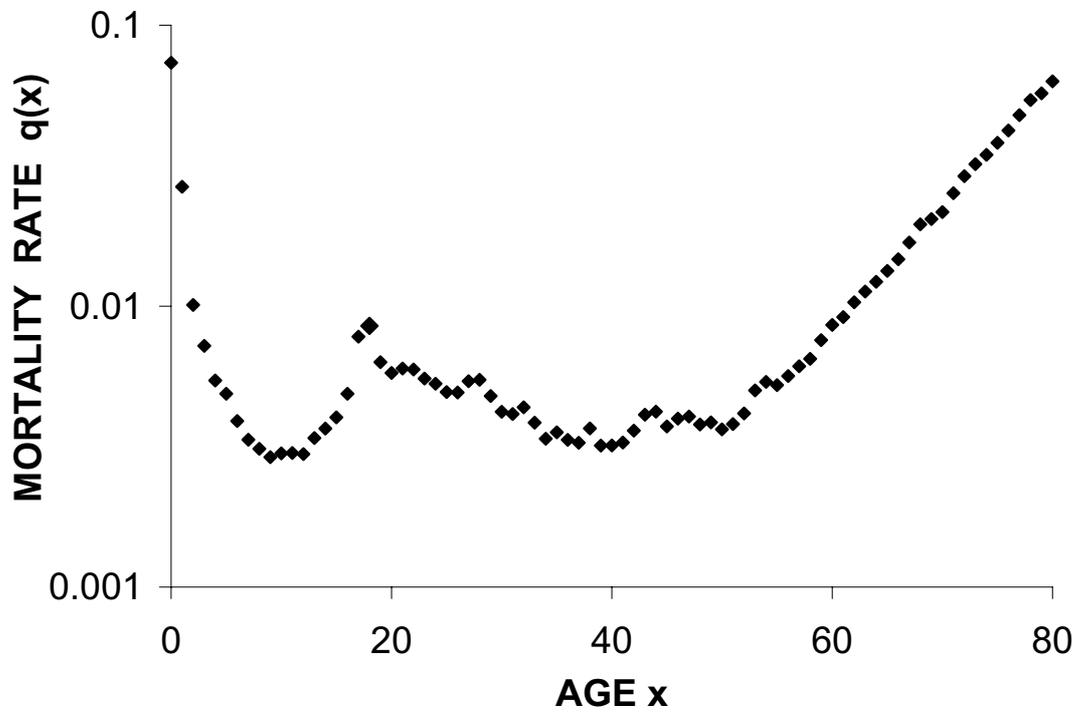

**Figure 4**

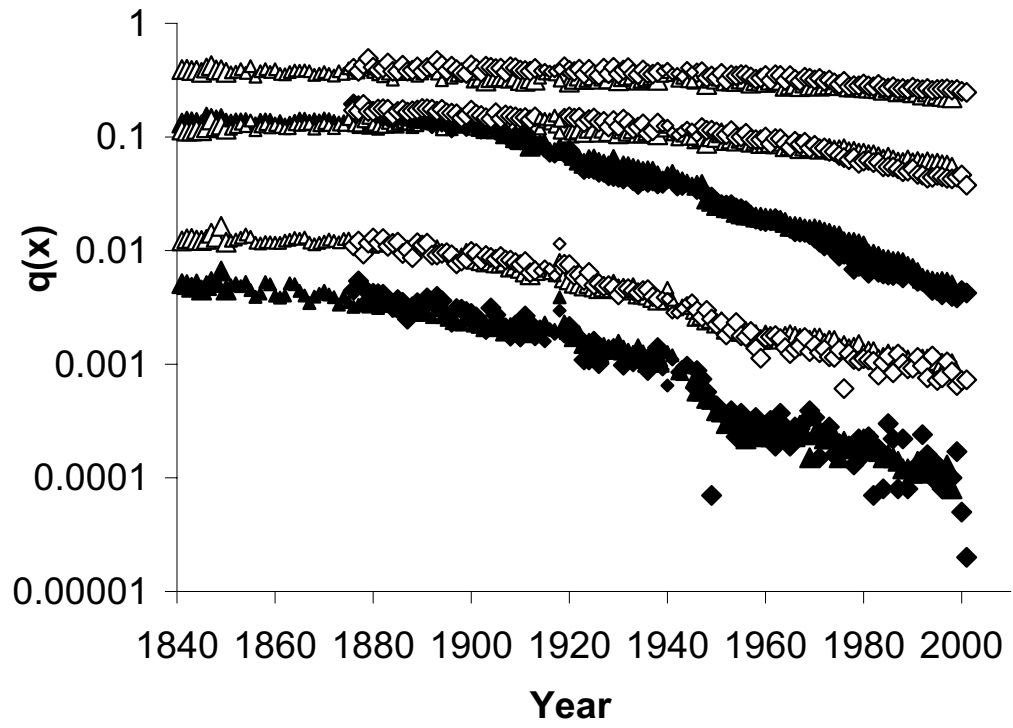

**Figure 5**

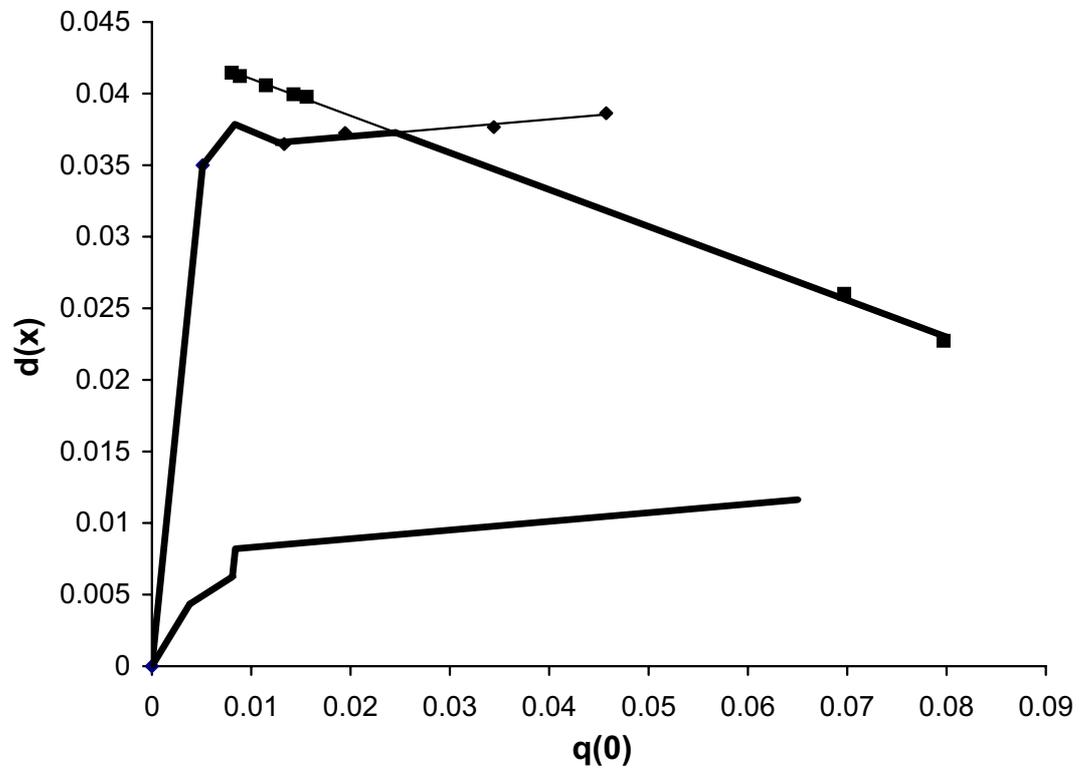

**Figure 6**

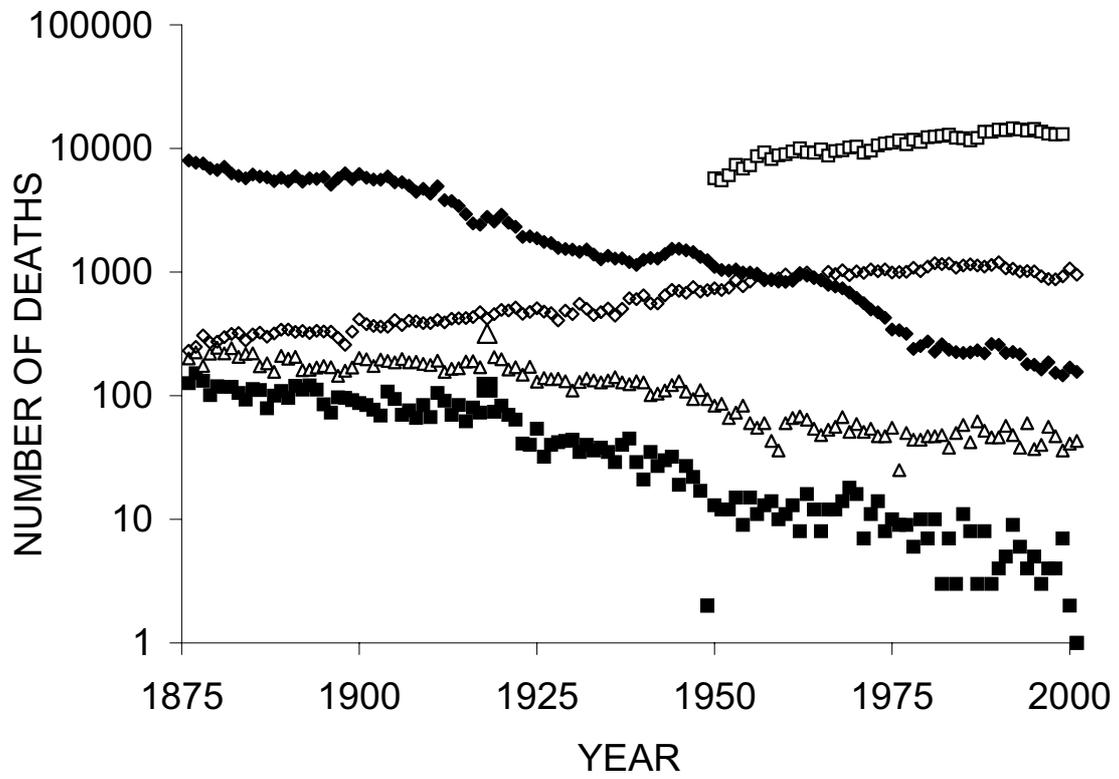